\input harvmac.tex
\input epsf


\def\figin{\epsfcheck\figin}\def\figins{\epsfcheck\figins}
\def\epsfcheck{\ifx\epsfbox\UnDeFiNeD
\message{(NO epsf.tex, FIGURES WILL BE IGNORED)}
\gdef\figin##1{\vskip2in}\gdef\figins##1{\hskip.5in}
\else\message{(FIGURES WILL BE INCLUDED)}%
\gdef\figin##1{##1}\gdef\figins##1{##1}\fi}
\def\DefWarn#1{}
\def\figinsert{\goodbreak\midinsert}
\def\ifig#1#2#3{\DefWarn#1\xdef#1{fig.~\the\figno}
\writedef{#1\leftbracket fig.\noexpand~\the\figno}%
\figinsert\figin{\centerline{#3}}\medskip\centerline{\vbox{\baselineskip12pt
\advance\hsize by -1truein\noindent\footnotefont{\bf Fig.~\the\figno:} #2}}
\bigskip\endinsert\global\advance\figno by1}



\def\unlockat{\catcode`\@=11}
\def\lockat{\catcode`\@=12}

\unlockat

\def\newsec#1{\global\advance\secno by1\message{(\the\secno. #1)}
\global\subsecno=0\global\subsubsecno=0\eqnres@t\noindent
{\bf\the\secno. #1}
\writetoca{{\secsym} {#1}}\par\nobreak\medskip\nobreak}
\global\newcount\subsecno \global\subsecno=0
\def\subsec#1{\global\advance\subsecno
by1\message{(\secsym\the\subsecno. #1)}
\ifnum\lastpenalty>9000\else\bigbreak\fi\global\subsubsecno=0
\noindent{\it\secsym\the\subsecno. #1}
\writetoca{\string\quad {\secsym\the\subsecno.} {#1}}
\par\nobreak\medskip\nobreak}
\global\newcount\subsubsecno \global\subsubsecno=0
\def\subsubsec#1{\global\advance\subsubsecno by1
\message{(\secsym\the\subsecno.\the\subsubsecno. #1)}
\ifnum\lastpenalty>9000\else\bigbreak\fi
\noindent\quad{\secsym\the\subsecno.\the\subsubsecno.}{#1}
\writetoca{\string\qquad{\secsym\the\subsecno.\the\subsubsecno.}{#1}}
\par\nobreak\medskip\nobreak}

\def\subsubseclab#1{\DefWarn#1\xdef
#1{\noexpand\hyperref{}{subsubsection}%
{\secsym\the\subsecno.\the\subsubsecno}%
{\secsym\the\subsecno.\the\subsubsecno}}%
\writedef{#1\leftbracket#1}\wrlabeL{#1=#1}}
\lockat

\def\IL{\relax{\rm I\kern-.18em L}}
\def\IH{\relax{\rm I\kern-.18em H}}
\def\IR{\relax{\rm I\kern-.18em R}}
\def\IC{\relax\hbox{$\inbar\kern-.3em{\rm C}$}}
\def\IZ{\relax\ifmmode\mathchoice
{\hbox{\cmss Z\kern-.4em Z}}{\hbox{\cmss Z\kern-.4em Z}}
{\lower.9pt\hbox{\cmsss Z\kern-.4em Z}}
{\lower1.2pt\hbox{\cmsss Z\kern-.4em Z}}\else{\cmss Z\kern-.4em
Z}\fi}
\def\CM {{\cal M}}
\def\CN {{\cal N}}

\def\CH {{\cal H}}
\def\CC {{\cal C}}
\def\CB {{\cal B}}

\def\CM {{\cal M}}
\def\CN {{\cal N}}

\def\cmtld{\widetilde{\CM} }

\font\manual=manfnt \def\dbend{\lower3.5pt\hbox{\manual\char127}}

\def\IZ{\relax\ifmmode\mathchoice
{\hbox{\cmss Z\kern-.4em Z}}{\hbox{\cmss Z\kern-.4em Z}}
{\lower.9pt\hbox{\cmsss Z\kern-.4em Z}}
{\lower1.2pt\hbox{\cmsss Z\kern-.4em Z}}\else{\cmss Z\kern-.4em
Z}\fi}
\def\half {{1\over 2}}

\def\lsim{
{\ \lower-1.2pt
\vbox{\hbox{\rlap{$>$}\lower5pt\vbox{\hbox{$\sim$}}}}\ }
}
\def\grsim{
{\ \lower-1.2pt\vbox{\hbox{\rlap{$<$}\lower5pt\vbox{\hbox{$\sim$}}}}\ }
}

\def\CM {{\cal M}}
\def\CN {{\cal N}}


\def\IZ{\relax\ifmmode\mathchoice
{\hbox{\cmss Z\kern-.4em Z}}{\hbox{\cmss Z\kern-.4em Z}}
{\lower.9pt\hbox{\cmsss Z\kern-.4em Z}}
{\lower1.2pt\hbox{\cmsss Z\kern-.4em Z}}\else{\cmss Z\kern-.4em
Z}\fi}
\def\IB{\relax{\rm I\kern-.18em B}}
\def\IC{{\relax\hbox{$\inbar\kern-.3em{\rm C}$}}}
\def\ID{\relax{\rm I\kern-.18em D}}
\def\IE{\relax{\rm I\kern-.18em E}}
\def\IF{\relax{\rm I\kern-.18em F}}
\def\IG{\relax\hbox{$\inbar\kern-.3em{\rm G}$}}
\def\IGa{\relax\hbox{${\rm I}\kern-.18em\Gamma$}}
\def\IH{\relax{\rm I\kern-.18em H}}
\def\II{\relax{\rm I\kern-.18em I}}
\def\IK{\relax{\rm I\kern-.18em K}}
\def\IP{\relax{\rm I\kern-.18em P}}

\def\IQ{\relax\hbox{$\inbar\kern-.3em{\rm Q}$}}
\def\IP{\relax{\rm I\kern-.18em P}}

\def\inbar{\,\vrule height1.5ex width.4pt depth0pt}

\def\mod{{\rm mod}}

\font\cmss=cmss10 \font\cmsss=cmss10 at 7pt
\def\IR{\relax{\rm I\kern-.18em R}}

\def\End{ {\rm End}} 
\def\Im{{\rm Im}}


\def\boxit#1{\vbox{\hrule\hbox{\vrule\kern8pt
\vbox{\hbox{\kern8pt}\hbox{\vbox{#1}}\hbox{\kern8pt}}
\kern8pt\vrule}\hrule}}
\def\mathboxit#1{\vbox{\hrule\hbox{\vrule\kern8pt\vbox{\kern8pt
\hbox{$\displaystyle #1$}\kern8pt}\kern8pt\vrule}\hrule}}


\def\inbar{\,\vrule height1.5ex width.4pt depth0pt}
\def\ndt{\noindent}

\font\cmss=cmss10 \font\cmsss=cmss10 at 7pt
\def\IR{\relax{\rm I\kern-.18em R}}

%


\lref\andermoore{G. Anderson and G. Moore, 
``Rationality in conformal field theory,'' 
Comm. Math. Phys. {\bf 117}(1988)441}

\lref\adf{L. Andrianopoli, R. D'Auria, and S. Ferrara,
``U-duality and central charges in various dimensions 
revisited,'' hep-th/9612105; ``Flat symplectic bundles 
of N-extended supergravities, central charges and 
black hole entropy,'' hep-th/9707203} 

\lref\adfii{L. Andrianopoli, R. D'Auria, and S. Ferrara,
``Duality, Central Charges and Entropy of Extremal BPS Black Holes,''
hep-th/9709113;
``Five Dimensional U-Duality, Black-Hole Entropy and Topological Invariants,''
hep-th/9705024, Phys.Lett. B411 (1997) 39-45;
``U-Duality, attractors and Bekenstein-Hawking
entropy in four and five dimensional supergravities,''
hep-th/?????;
``U-Invariants, Black-Hole Entropy and Fixed Scalars,''
hep-th/9703156,Phys.Lett. B403 (1997) 12-19;
``Central Extension of Extended Supergravities in Diverse Dimensions,''
hep-th/9608015, Int.J.Mod.Phys. A12 (1997) 3759-3774}

\lref\adfesev{L. Andrianopoli, R. D'Auria,  S. Ferrara,  P. Fre',  M.
Trigiante,
``$E_{7(7)}$ Duality, BPS Black-Hole Evolution and Fixed Scalars,''
hep-th/9707087}

\lref\enntwosugra{
L. Andrianopoli,  M. Bertolini,  A. Ceresole,  R. D'Auria,  S. Ferrara,  P.
Fre', 
T. Magri, ``N=2 Supergravity and N=2 Super Yang-Mills Theory on General Scalar
Manifolds: Symplectic Covariance, Gaugings and the Momentum Map,''
hep-th/9605032,J.Geom.Phys. 23 (1997) 111-189 ; 
``General Matter Coupled N=2 Supergravity,'' hep-th/9603004, 
Nucl.Phys. B476 (1996) 397-417} 

\lref\adfl{L. Andrianopoli, R. D'Auria, S. Ferrara, M.A. Lledo, 
``Horizon geometry, duality and fixed scalars in six 
dimensions,'' hep-th/9802147}  

\lref\asplut{P. Aspinwall and C. Lutkin,  ``Geometry of
mirror manifolds,'' Nucl. Phys. {\bf B355}(1991)482}

\lref\agm{P. Aspinwall, B. Greene, and D. Morrison,
``Calabi-Yau Moduli Space, Mirror Manifolds and Spacetime Topology Change in
String Theory,''  hep-th/9309097;
``Measuring Small Distances in N=2 Sigma Models,''
hep-th/9311042}

\lref\aspmorr{P. Aspinwall and D. Morrison, 
``String Theory on K3 Surfaces,''  hep-th/9404151}

\lref\aspinwall{P. Aspinwall, ``K3 surfaces and
string duality,'' hep-th/9611137} 
 
\lref\amiii{P. Aspinwall and D. Morrison,
``Point-like Instantons on K3 Orbifolds,'' 
hep-th/9705104, Nucl.Phys. B503 (1997) 533-564}

\lref\asphm{P. Aspinwall, ``Aspects of the hypermultiplet
moduli space in string duality,'' hep-th/9802194.}

\lref\amrp{P. Aspinwall and D. Morrison,
``Non-simply-connected gauge groups and rational
points on elliptic curves,'' hep-th/9805206}

\lref\balasub{V.J. Balasubramanian,
``How To Count the States of Extremal Black Holes in N=8 Supergravity,''
hep-th/9712215}

\lref\bpv{W. Barth, C. Peters, A. Van de Ven, 
{\it Compact Complex Surfaces}, Springer-Verlag 1984}

\lref\batyrev{V.V. Batyrev, ``Variations of the mixed Hodge
structure of affine hypersurfaces in algebraic tori,''
Duke Math. J. {\bf 69}(1993) 349; ``Dual polyhedra and
mirror symmetry for Calabi-Yau hypersurfaces in toric varieties,'' J. Algebraic
Geom. {\bf 3}(1994) 493}

\lref\bbs{K. Becker, M. Becker, and A. Strominger, 
``Fivebranes, Membranes and Non-Perturbative String Theory,''
hep-th/9507158;Nucl.Phys. B456 (1995) 130-152}

\lref\cardi{ K. Behrndt, G. Lopes Cardoso, B. de Wit, R. Kallosh, D. Löst, T.
Mohaupt, ``Classical and quantum N=2 supersymmetric black holes,'' 
hep-th/9610105} 

\lref\cardii{K. Behrndt, G. Lopes Cardoso, I. Gaida, 
``Quantum N = 2 Supersymmetric Black Holes in the S-T Model,''
hep-th/9704095; K. Behrndt and I. Gaida,
``Subleading contributions from instanton corrections in
N=2 supersymmetric black hole entropy,'' hep-th/9702168} 

\lref\bls{K. Behrndt,  D. L\"ust, W.A.  Sabra, 
``Stationary solutions of N=2 supergravity,'' 
hep-th/9705169,Nucl.Phys. B510 (1998) 264-288}

\lref\berglund{P. Berglund, P. Candelas, X. de la Ossa, 
A. Font, T. Hubsch, D. Jancic, and F. Quevedo, 
``Periods for Calabi-Yau and Landau-Ginzburg vacua,'' 
hep-th/9308005}

\lref\bkmt{P. Berglund, A. Klemm, P. Mayr, and
S. Theisen, ``On type IIB vacua with varying
coupling constant,'' hep-th/9805189}

\lref\berpan{M. Bershadsky, T. Pantev, and V. Sadov,
``F-Theory with quantized fluxes,'' hep-th/9805056}

\lref\birklang{H. Birkenhake and Ch. Lange, 
{\it Complex Abelian Varieties}, Springer Verlag 1992}

\lref\borcea{C. Borcea, ``Calabi-Yau threefolds and
complex multiplication,'' in {\it Essays on Mirror
Manifolds}, S.-T. Yau, ed., International Press, 1992.} 

\lref\borcha{R. E. Borcherds, ``The monster Lie algebra,'' Adv. Math. {\bf 83}
No. 1 (1990).}

\lref\borchi{R. Borcherds,``Monstrous moonshine
and monstrous Lie superalgebras,'' Invent. Math.
{\bf 109}(1992) 405.}

\lref\borchii{R. Borcherds, ``Automorphic forms
on $O_{s+2,2}(R)$ and infinite products,''
Invent. Math. {\bf 120}(1995) 161.}

\lref\borchiii{R. Borcherds, ``Automorphic forms
on $O_{s+2,2}(R)^+$ and generalized Kac-Moody
algebras,'' contribution to the Proceedings of
the 1994 ICM, Zurich. }

\lref\borchiv{R. Borcherds, ``The moduli space
of Enriques surfaces and the fake monster Lie
superalgebra,''  preprint (1994).}

\lref\borchalg{R. Borcherds, ``Generalized Kac-Moody algebras,'' Journal of
Algebra
{\bf 115} (1988) 501.}

\lref\borevich{Z.I. Borevich and I.R. Shafarevich, 
{\it Number Theory} Academic Press 1966}

\lref\buell{D. Buell, {\it Binary quadratic forms} 
Springer-Verlag, 1989}

\lref\cdgp{P. Candelas, X. de la Ossa, P. S. Green and L. Parkes,
``A pair of Calabi-Yau manifolds as an exactly soluble
superconformal theory,'' Nucl. Phys. {\bf B359} (1991) 21.}

\lref\twop{P. Candelas X. de la Ossa, A. Font, 
S. Katz, and D.R. Morrison, ``Mirror symmetry 
for two-parameter models - I,'' 
hep-th/9308083}

\lref\twopii{ P. Candelas, A. Font, 
S. Katz, and D.R. Morrison,
``Mirror Symmetry for Two Parameter Models -- II,''
hep-th/9403187}

\lref\cardcurl{G. L. Cardoso, G. Curio,   
D. L\"{u}st, and T. Mohaupt, ``On the Duality between the Heterotic String and
F-Theory in 8 Dimensions,''  hep-th/9609111} 

\lref\clm{G. L. Cardoso, D. L\"{u}st and T. Mohaupt,
``Threshold corrections and symmetry enhancement in string
compactifications,'' Nucl. Phys. {\bf B450} (1995) 115,
hep-th/9412209.}

\lref\lust{Gabriel Lopes Cardoso, Gottfried Curio, Dieter Lust, Thomas Mohaupt,
``On the Duality between the Heterotic String and F-Theory in 8 Dimensions,''
hep-th/9609111}

\lref\cassels{J.W.S. Cassels, {\it Rational Quadratic Forms}, 
Academic Press, 1978} 

\lref\cecotti{S. Cecotti, ``$N=2$ supergravity, type IIB 
superstrings, and algebraic geometry,'' 
Commun.Math.Phys.131:517-536,1990} 

\lref\cdfvp{A. Ceresole,  R. D'Auria,  S. Ferrara,  A. Van Proeyen,
``Duality Transformations in Supersymmetric Yang-Mills Theories coupled to
Supergravity,''  hep-th/9502072}

\lref\fvdeei{A. Chamseddine, S. Ferrara, G. Gibbons,
and R. Kallosh, ``Enhancement of Supersymmetry Near 5d Black Hole Horizon,''
hep-th/9610155}

\lref\fvdeeii{A. Chou,  R. Kallosh,  J. Rahmfeld, S.-J. Rey,  M. Shmakova,
W.K. Wong,
``Critical Points and Phase Transitions in 5D Compactifications of M-Theory,''
hep-th/9704142}

\lref\clemens{H. Clemens, {\it A scrapbook of
complex curve theory}  Plenum Press, 1980
}

\lref\cohen{H. Cohen,
{\it A Course in Computational Algebraic Number Theory}, Springer GTM}

\lref\pbcohen{P.B. Cohen, ``Humbert surfaces and 
transcendence properties of automorphic 
functions,'' Rocky Mountain J. Math. {\bf 26}
(1996) 987}

\lref\slg{J.H. Conway and N.J.A. Sloane, 
{\it Sphere Packings, Lattices, and Codes}, 
Springer-Verlag, 1993 } 

\lref\csiv{J.H. Conway and N.J.A. Sloane,
``Low-dimensional lattices. IV. The mass
formula,'' Proc. R. Soc. Lond. {\bf A419} (1988)259.}

\lref\cox{D.A. Cox, {\it Primes of the form $x^2 + n y^2$}, John Wiley, 1989.}

\lref\cj{E. Cremmer and B. Julia, ``The $SO(8)$ supergravity''
Nuc. Phys. {\bf B159}(1979)141}

\lref\cvetic{M. Cvetic, A. A. Tseytlin,
``Solitonic Strings and BPS Saturated Dyonic Black Holes,''
hep-th/9512031; M. Cvetic and D. Youm, ``All the static sperically
symmetric black holes of heterotic string on a six torus,''
hep-th/9512127}

\lref\deligne{P. Deligne, ``La conjecture de Weil pour 
les surfaces K3,'' Inv. Math. {\bf 15} (1972) 206} 

\lref\delignelcsl{P. Deligne, ``Local behavior of Hodge structures at
infinity,'' in {\it Mirror Symmetry II}, B. Greene and 
S.-T. Yau eds. International Press 1991.} 

\lref\deser{S. Deser, A. Gomberoff, M. Henneaux, 
and C. Teitelboim, ``p-Brane Dyons and 
Electric-magnetic Duality,'' hep-th/9712189}

\lref\difrancesco{P. Di Francesco,  P. Mathieu, and
D. S\'en\'echal, {\it Conformal Field Theory} Springer 1997}

\lref\dv{R. Dijkgraaf and E. Verlinde, ``Modular invariance and
the fusion algebras,'' Nucl. Phys. (Proc. Suppl) {\bf 5B} (1988)
110}

\lref\dolgachev{I. Dolgachev,
``Mirror symmetry for lattice polarized K3 surfaces,''
alg-geom/9502005}

\lref\dolgachevi{I. Dolgachev, ``Integral quadratic forms: 
applications to algebraic geometry,'' Sem. Bourbaki, 
1982, no. 611, p. 251} 

\lref\donagi{R. Donagi, ``ICMP lecture on Heterotic/F-theory
duality,'' hep-th/9802093}

\lref\fhsv{
S. Ferrara, J. A. Harvey, A. Strominger, C. Vafa ,
``Second-Quantized Mirror Symmetry, '' Phys. Lett. {\bf B361} (1995) 59;
hep-th/9505162. }

\lref\fgk{S. Ferrara,  G. W. Gibbons,  R. Kallosh,
``Black Holes and Critical Points in Moduli Space,''  hep-th/9702103}

\lref\fks{S. Ferrara, R. Kallosh, and A. Strominger,
``N=2 Extremal Black Holes,''   hep-th/9508072}

\lref\fk{S. Ferrara and R. Kallosh, ``Universality of Sypersymmetric
Attractors,''   hep-th/9603090;  ``Supersymmetry and Attractors,''  
hep-th/9602136; S. Ferrara, ``Bertotti-Robinson Geometry and Supersymmetry,''
hep-th/9701163}

\lref\fklz{S. Ferrara, C. Kounnas, D. L\"{u}st and F. Zwirner,
``Duality-invariant
partition functions and automorphic superpotentials for $(2,2)$ string
compactifications,''  Nucl. Phys.  {\bf B365} (1991) 431. }

\lref\fm{S. Ferrara and J.M. Maldacena,
``Branes, central charges and U-duality invariant BPS conditions,''  
hep-th/9706097}

\lref\fg{S. Ferrara and M. G\"unaydin,
``Orbits of Exceptional Groups, Duality and BPS States in String Theory,''
hep-th/9708025}

\lref\font{A. Font, ``Periods and duality symmetries in
Calabi-Yau compactifications,'' hep-th/9203084}

\lref\fre{P. Fr\'e, ``Supersymmetry and First Order Equations for Extremal
States: Monopoles, Hyperinstantons, Black-Holes and p-Branes,'' 
hep-th/9701054,Nucl.Phys.Proc.Suppl. 57 (1997) 52-64 } 

\lref\fs{D. Friedan and S. Shenker, ``The integrable analytic 
geometry of quantum string,'' Phys. Lett. {\bf 175B}(1986) 287;
``The analytic geometry of two dimensional conformal 
field theory,'' Nucl. Phys. {\bf B281}(1987) 509; 
D. Friedan, ``The space of conformal field theories and the 
space of classical string ground states,'' in 
{\it Physics and Mathematics of Strings}, L. Brink, D. Friedan, 
and A.M. Polyakov eds., World Scientific 1990}

\lref\fmw{R. Friedman, J. Morgan, and E. Witten, 
``Vector Bundles And F Theory,'' hep-th/9701162,
Commun.Math.Phys. 187 (1997) 679-743; 
``Principal G-bundles over elliptic curves,'' 
alg-geom/9707004} 

\lref\fvdeeiv{I. Gaida,
``N = 2 Supersymmetric Quantum Black Holes in Five Dimensional Heterotic String
Vacua,''
hep-th/980214}

\lref\gms{O. Ganor, D. Morrison, and N. Seiberg,}

\lref\gauss{C.F. Gauss, {\it Disquisitiones Arithmeticae}, 
Leipzig, 1801. English translation, Yale University 
Press, 1966. } 

\lref\gordon{B. Brent Gordon, ``A survey of the
Hodge conjecture for Abelian varieties,''
alg-geom/????}

\lref\bhgross{B.H. Gross, ``Groups over $\IZ$,'' 
Invent. Math. {\bf 124}(1996) 263} 

\lref\gzsm{B. Gross and D. Zagier,
``On singular moduli,'' J. reine angew. Math.
{\bf 355} (1985) 191}

\lref\greenekantor{B.R. Greene and 
Y. Kanter, ``Small Volumes in Compactified String Theory,''
hep-th/9612181}

\lref\griffiths{P. Griffiths, et. al.
{\it Topics in transcendental algebraic geometry}, Ann. Math. Studies.
Princeton Univ. 
Press, Princeton, 1984} 

\lref\gkp{S. Gubser, I. Klebanov, and A. Polykov,
``Gauge Theory Correlators from Non-Critical String Theory,''
hep-th/9802109}

\lref\harris{J. Harris, {\it Algebraic Geometry} Springer 1992}

\lref\hmi{ J. Harvey and G. Moore, ``Algebras, BPS States, and Strings,''
hep-th/9510182;
Nucl.Phys. B463 (1996) 315-368}

\lref\hmalg{J. Harvey and G. Moore,
``On the algebras of BPS states,'' 
hep-th/9609017}

\lref\hilbert{D. Hilbert, ``Mathematical problems,''
in {\it Mathematical developments arising from
Hilbert problems}, Proc. Symp. Pure. Math.
{\bf 28} vol. 1. AMS 1976}

\lref\hornemoore{J. Horne and G. Moore } 

\lref\hulekgrit{K. Hulek and V. Gritsenko}

\lref\hull{C. Hull and P. Townsend, ``Unity of superstring dualities,''  hep-th/9410167 }

\lref\husemoller{D. Husemoller, Elliptic curves}

\lref\hkty{ S. Hosono, A. Klemm, S. Theisen and S.-T. Yau,
``Mirror Symmetry, Mirror Map and Applications to Complete Intersection
Calabi-Yau Spaces, ''
hep-th/9406055;Nucl.Phys. B433 (1995) 501-554}

\lref\hkt{S. Hosono, A. Klemm, and S. Theisen,
``Lectures on Mirror Symmetry,'' hep-th/????}

\lref\itzykson{{\it From Number Theory to 
Physics}, C. Itzykson, J.-M. Luck, P. Moussa, 
and M. Waldschmidt, eds., Springer Verlag, 
1992}

\lref\ireland{K. Ireland and M. Rosen, {\it A Classical
Introduction to Modern Number Theory},
Springer 1990}

\lref\kalkol{R. Kallosh and B. Kol,
``E(7) Symmetric Area of the Black Hole Horizon,'' 
hep-th/9602014} 

\lref\klm{A. Klemm, W. Lerch, and P. Mayr, }

\lref\klemmtheis{A. Klemm and S. Theisen, 
``Considerations of One-Modulus Calabi-Yau 
Compactifications: Picard-Fuchs equations, 
Kahler potentials and mirror maps,'' 
hep-th/9205041}

\lref\hkmiii{R. Kobayashi and A.N. Todorov, 
``Polarized Period Map for Generalized K3 Surfaces and 
the Moduli of Einstein Metrics,'' Tohoku Math. J. 
{\bf 39} (1987) 341} 

\lref\langi{S. Lang, {\it Algebra}, Addison-Wesley, 1971} 

\lref\langii{S. Lang, {\it Complex Multiplication}
Springer  1983}

\lref\langdg{S. Lang, {\it Survey of Diophantine
Geometry}, Springer Verlag 1997}

\lref\langlands{R.P. Langlands,
``Some contemporary problems with
origins in the Jugendtraum,''
in \hilbert.}

\lref\llw{W. Lerche, D. L\"ust, and N. Warner, 
Phys. Lett. {\bf 231B}(1989)417}

\lref\lerchstie{W. Lerche and S. Steiberger,
``Prepotential, Mirror Map, and F-theory on
K3,'' hep-th/9804176}

\lref\lianyau{B.H. Lian and S.T. Yau, 
``Arithmetic Properties of Mirror Map and Quantum Coupling,''
hep-th/9411234; Commun.Math.Phys. 176 (1996) 163-192;
``Mirror Maps, Modular Relations and Hypergeometric Series I,
hep-th/9507151}

\lref\luck{{\it Number Theory and Physics}, 
J.-M. Luck, P. Moussa, and M. Waldschmidt, eds. 
Proceedings in Physics {\bf 47}, Springer Verlag 
1990}

\lref\lustrev{D. L\"ust, ``String Vacua with 
N=2 Supersymmetry in Four Dimensions,''
hep-th/9803072} 

\lref\msw{J. Maldacena, A. Strominger, and E. Witten,
``Black Hole Entropy in M-Theory,''  hep-th/9711053} 

\lref\maldacena{J. Maldacena,``The Large N Limit of Superconformal Field
Theories and Supergravity.''  hep-th/9711200} 

\lref\mazur{B. Mazur, ``Arithmetic on curves,''
Bull. AMS {\bf 14}(1986) 207}

\lref\mikhailov{A. Mikhailov, ``Momentum Lattice for CHL String,''
hep-th/9806030 }

\lref\mirandai{R. Miranda and D.R. Morrison,
``The number of embeddings of integral quadratic forms. I,II''
Proc. Japan Acad. {\bf 61}(1985) 217}

\lref\ms{G. Moore and N. Seiberg,
``Naturality in conformal field theory,''
Nucl. Phys. {\bf B313}(1989)16 }

\lref\aaintro{G. Moore, ``Attractors and Arithmetic,'' 
hep-th/????} 

\lref\aalong{G. Moore, ``Arithmetic and Attractors,'' 
hep-th/9807087} 

\lref\hkmii{D.R. Morrison, ``Some remarks on the 
moduli of K3 surfaces,'' in K. Ueno ed. 
{\it Classification of Algebraic and Analytic 
Manifolds}, Prog. in Math. {\bf 39}, 
Birkh\"auser, 1983}

\lref\morrpicard{D.R. Morrison,
``On $K3$ surfaces with large Picard number,''
Invent..Math.75 (1984), no. 1, 105--121}

\lref\dmlcsl{D.R. Morrison, ``Making enumerative 
predictions by means of mirror symmetry,'' 
  in {\it Mirror Symmetry II}, B. Greene and 
S.-T. Yau eds. International Press 1991.} 

\lref\dmcomps{D.R. Morrison,
``Compactifications of moduli spaces inspired by mirror symmetry,''
alg-geom/9304007}

\lref\lookingglass{D.R. Morrison, ``Through the 
Looking Glass,'' alg-geom/9705028}

\lref\vfmr{D. Morrison and C. Vafa, ``Compactifications of F-theory on
Calabi-Yau threefolds -I,'' hep-th/9602114;``Compactifications of F-theory on

Calabi-Yau threefolds -II,'' hep-th/9603161}

\lref\mumford{D. Mumford, {\it Tata Lectures on
Theta I}, Birkh\"auser 1983}

\lref\nikulin{V. Nikulin, ``Integral symmetric
bilinear forms and some of their applications,''
Math. Izv. {\bf 14} (1980) 103}

\lref\swi{N. Seiberg and E. Witten,
``Monopole Condensation, And Confinement In $N=2$ Supersymmetric Yang-Mills
Theory,''
hep-th/9407087;Nucl. Phys. {\bf B426} (1994) 19.}

\lref\swii{N. Seiberg and E. Witten,
``Monopoles, Duality and Chiral Symmetry Breaking in N=2 Supersymmetric QCD,''
hep-th/9408099;Nucl. Phys. {\bf B431} (1994) 484.}

\lref\shiodaef{T. Shioda, ``On elliptic modular surfaces,''
J. Math. Soc. Japan {\bf 24}(1972) 20}

\lref\shiodamitani{T. Shioda and N. Mitani,
``Singular abelian surfaces and binary quadratic
forms,'' in {\it Classification of algebraic varieties and comopact complex
manifolds} SLN 412 (1974) 259.}

\lref\shioda{T. Shioda and H. Inose, ``On singular
K3 surfaces,'' in Complex analysis and algebraic geometry,
Cambridge University Press, Cambridge, 1977}

\lref\stark{H. Stark, in {\it From Number Theory
to Physics} C. Itzykson et. al. eds FIX}

\lref\psshaf{I.I. Piatetski-Shapiro and I.R. Shafarevich, 
``A Torelli theorem for algebraic surfaces of type 
$K3$,'' Math. USSR Izvestia {\bf 5}(1971)547}

\lref\sabra{W.A. Sabra, ``Black holes in N=2 supergravity theories and harmonic
functions,''  hep-th/9704147, Nucl.Phys. B510 (1998) 247-263}

\lref\fvdeeiii{W. Sabra,
``General BPS Black Holes In Five Dimensions,''
hep-th/9708103}

\lref\schmakova{M. Schmakova, ``Calabi-Yau black holes,'' 
hep-th/9612076}

\lref\serre{J.-P. Serre, {\it Statement of results},
in {\it Seminar on Complex Multiplication},
A. Borel et. al. eds., SLN 21, 1966}

\lref\shiga{H. Shiga and J. Wolfart, 
``Criteria for complex multiplication and 
transcendence properties of automorphic 
functions,'' J. reine angew. Math. {\bf 463} (1995) 1}

\lref\shimtan{G. Shimura and Y. Taniyama,
{\it Complex Multiplication of Abelian Varieties
and its applications to number theory}, Math. Soc. of Japan,
1961.}

\lref\gshimura{G. Shimura, {\it Abelian
Varieties with Complex Multiplication and Modular Functions}, Princeton
University Press, Princeton, 1998} 

\lref\silvermanag{J. Silverman, {\it Arithmetic Geometry},
G. Cornell and J.H. Silverman, eds. Springer Verlag 1986}

\lref\silveradvtop{J. Silverman, {\it Advanced Topics in the
Arithmetic of Elliptic Curves} Springer Verlag GTM 151, 1994}

\lref\conifold{A. Strominger, ``Massless Black Holes
and Conifolds in String Theory
,'' hep-th/9504090 } 

\lref\sv{A. Strominger and C. Vafa, 
``Microscopic Origin of the Bekenstein-Hawking Entropy,''
hep-th/9601029; Phys.Lett. B379 (1996) 99-104 }

\lref\stromi{A. Strominger, ``Macroscopic Entropy of $N=2$ Extremal Black
Holes,''  hep-th/9602111}

\lref\syz{A. Strominger, S.-T. Yau, and E. Zaslow,
``Mirror symmetry is T-duality,'' hep-th/9606040,
Nucl.Phys. B479 (1996) 243-259}

\lref\takeuchi{K. Takeuchi, ``Arithmetic triangle
groups,'' J. Math. Soc. Japan {\bf 29} (1977) 91}

\lref\taormina{A. Taormina and S.M.J. Wilson,
``Virasoro character identities and Artin L-functions,''
physics/9706004}

\lref\thomas{R. Thomas, ``A holomorphic Casson
invariant for Calabi-Yau 3-folds and bundles on
K3 fibrations,'' IAS preprint}

\lref\hkmi{A. Todorov, ``Applications of K\"ahler-Einstein-
Calabi-Yau Metric to Moduli of K3 surfaces,'' 
Inv. Math. {\bf 61} (1980) 251} 

\lref\ueno{K. Ueno, ``On fibre spaces of normally 
polarized abelian varieties of dimension 2,'' J. Fac. Sci. 
Tokyo, {\bf 18}(1971) 37}

\lref\vafa{C. Vafa, ``Evidence for F-theory,'' hep-th/9602022}

\lref\vafacyi{C. Vafa, ``Black Holes and Calabi-Yau Threefolds,''
hep-th/9711067}

\lref\vafacyii{C. Vafa, ``Extending Mirror Conjecture to Calabi-Yau with
Bundles,''
hep-th/9804131}

\lref\viehweg{Viehweg}

\lref\vladut{S.G. Vladut, {\it Kronecker's Jugendtraumand modular functions},
Gordon and Breach, 1991.}

\lref\weil{A. Weil, {\it Elliptic functions according
to Eisenstein and Kronecker}, Springer-Verlag 1976 } 

\lref\weilii{A. Weil, ``The field of definition of a
variety,'' Amer. J. Math. {\bf 78}(1956) 509}

\lref\grssmm{E. Witten, ``Quantum field theory, 
grassmannians and algebraic curves,'' Commun.Math.Phys.113:529,1988} 

\lref\WittVar{E. Witten, ``String theory in various dimensions,'' 
hep-th/9503124}

\lref\wittenholog{E. Witten, ``Anti De Sitter Space And Holography,''
hep-th/9802150}

\Title{\vbox{\baselineskip12pt
\hbox{hep-th/9807056}
\hbox{YCTP-P16-98    }
}}
{\vbox{\centerline{Attractors and Arithmetic}
 }}

\bigskip
\centerline{Gregory Moore}
\bigskip
\centerline{\sl Department of Physics, Yale University}
\centerline{\sl New Haven, CT  06511}
\centerline{ \it moore@castalia.physics.yale.edu }

\bigskip
\bigskip
\noindent
We consider attractor varieties arising in the 
construction of dyonic black holes in Calabi-Yau 
compactifications of IIB string theory. We show 
that the attractor varieties are constructed from 
products of elliptic curves with complex multiplication 
for $\CN=4,8$ compactifications. The heterotic 
dual theories are related to rational conformal field 
theories. The emergence of curves with complex 
multiplication suggests many interesting connections 
between arithmetic and string theory. This paper is 
a brief overview of a longer companion paper 
entitled ``Arithmetic and Attractors'' hep-th/9807087.

\Date{July 2, 1998}

\newsec{Introduction}

This paper is a short introduction to 
\aalong. We have written it as a separate 
note with the hope that the present paper is 
something that people might actually read. 
Further explanations, more precise statements, 
other results, 
and more complete references can be found in \aalong. 

\newsec{The attractor equations}

The attractor mechanism, discovered by 
Ferrara, Kallosh and Strominger in \fks,
is a fascinating phenomenon that combines 
supersymmetric black holes, dynamical 
systems, and, as we show 
here, number theory.  Put briefly, in 
constructing spherically symmetric 
dyonic black holes in $d=4,\CN=2$ 
supergravity coupled to abelian vectormultiplets, 
one finds that the radial 
evolution of vectormultiplet scalars is 
described by a dynamical system. Under good 
conditions, (leading to  a black hole 
with a smooth horizon of positive area),  
the vectormultiplets flow to a fixed point 
in their target space. The fixed point is 
determined by the dyonic charge $\hat\gamma$ 
of the black hole, where $\hat\gamma$ is a vector in 
the lattice $\Lambda$ of electric and magnetic charges 
of the $\CN=2$ abelian gauge theory. 
Suppose the  
supergravity is the low-energy limit 
of type IIB superstring theory compactified on 
a Calabi-Yau (CY) 3-fold $X$. Then,  the vectormultiplet 
moduli space is identified with $\CM$,  the 
moduli space of complex structures on $X$, and
the charge lattice is $\Lambda= H^3(X;\IZ)$. 
The equation for the fixed point complex 
structure, called the attractor equation
for the charge $\hat \gamma\in H^3(X;\IZ)$,
is a condition on the Hodge structure 
of $X$  \stromi\fk:
\eqn\hodgede{
\hat \gamma = \hat\gamma^{3,0} + \hat \gamma^{0,3}.
}
If solutions to \hodgede\ exist then they are isolated 
points in $\CM$. 

Equation \hodgede\ also turns out to be related to 
a {\it nonperturbative} statement about BPS states
\fgk.  Indeed, $\hat \gamma$ specifies a 
superselection sector $\CH_{\hat\gamma}$ in the Hilbert space 
of states for the IIB compactification on $X$. 
The central charge   of  
the supersymmetry algebra 
in the superselection sector $\CH_{\hat\gamma}$ 
may be expressed in terms of a nowhere 
vanishing holomorphic 
$(3,0)$ form $\Omega$ on $X$. If $\gamma$ is 
Poincar\'e dual to $\hat \gamma$ the
central charge $Z(\Omega;\gamma)$ satisfies: 
\eqn\bpsmass{
\vert Z(\Omega;\gamma) \vert^2 \equiv {\vert \int_\gamma \Omega \vert^2 \over
i \int \Omega \wedge \bar \Omega}  \quad .
}
The BPS states of charge $\hat \gamma$ 
are the single-particle states in 
$\CH_{\hat\gamma}$ saturating the Bogomolnyi
bound. They might or might not exist, depending 
on $\hat \gamma$. If $\gamma$  does support a BPS state then the 
mass of the BPS state in Planck units is just \bpsmass. 
Note that, because of monodromy, we must  
consider the target space of the 
vectormultiplet scalars to be in the 
universal cover $\widetilde{\CM}$. 
Note too that the mass depends only on the 
point $z\in \widetilde{\CM}$ in Teichm\"uller 
space and not on the choice of $\Omega$.
The relation between BPS masses and the 
attractor equation \hodgede\ follows from the: 

\bigskip
\ndt
{\bf Theorem} \fgk. 
$\vert Z(z;\gamma) \vert^2 $
has a  stationary point at $z= z_*(\gamma) \in \widetilde{\CM}$, 
with $Z(z_*;\gamma) \not=0$ iff $\hat \gamma$ 
has Hodge decomposition $\hat \gamma = \hat \gamma^{3,0} + \hat \gamma^{0,3}$. Moreover, if 
such a stationary point exists in the interior of 
$\cmtld$ then 
it is a local minimum of $\vert Z(z;\gamma) \vert^2 $.

This theorem is very important to our considerations 
because, while   supergravity is only an approximation 
at large charges, the minimization principle applies to
an exact formula for BPS states. 
Hence it is sensible to 
think about the exact nature of the attractor varieties 
\hodgede\  even for small charges. 

This theorem also raises the question of whether 
attractor points are {\it global} minima of the BPS 
mass \bpsmass. In \aalong, section 9.2, we give an example 
of a compactification and charges $\gamma$ 
such that the equation  \hodgede\ has more than 
one distinct solution in $\widetilde{\CM}$ 
(not related by the duality  group.)  If our 
example is correct then it might present an interesting 
twist on the D-brane interpretation of black 
hole entropy \sv. In any case, the example shows 
that the the dynamical system 
of \fks\ defined by a charge $\gamma$ can have 
more than one basin of attraction $\CB$. The entropy 
is determined by the data $(\gamma,\CB)$. We 
refer to this data as an {\it area code}.

\newsec{Some solutions to the attractor equations}
 
In this section we describe solutions to the 
attractor equations for $d=4$ compactifications with 
$\CN=8,4$ supersymmetry. 

Let  $X$ be 
 a 3-dimensional complex torus. Compactification 
of $IIB$ superstrings on $X$ leads to $d=4, \CN=8$ 
supergravity. There are 28 independent
abelian gauge fields and 
the electric/magnetic charge lattice $\Lambda$ is a rank 56 
module for an integral form of the 
maximally split form of $E_7$, often called 
$E_{7,7}(\IZ)$ in the physics literature
\hull\WittVar. This form 
of $E_7$ preserves a certain quartic form $I_4(\gamma)$ 
on $\Lambda$ \cj\kalkol. Perturbative string 
theory and D-branes give a model for 
$\Lambda$ as the lattice 
$H^{\rm odd}(X;\IZ)\oplus II^{6,6}\oplus II^{6,6}$ 
where $II^{r,s}$ is the even unimodular lattice of signature 
$(-1)^r, (+1)^s$. In particular,    
$H^3(X;\IZ)$ is a submodule of $\Lambda$. 
By considerations of $R$-symmetry the 
 compactification moduli can be divided   into 
``vectormultiplets''  and ``hypermultiplets''  (in 
the terminology of $\CN=2$ representations)
and the attractor mechanism fixes the values 
of the vectormultiplets, leaving the hypermultiplets 
arbitrary \adfesev. 
Chief among 
 the vectormultiplets are the complex structure moduli. 
By $U$-duality we can take $\hat\gamma \in H^3(X;\IZ)$ 
and the  complex structure is then fixed by \hodgede.
The solution to the equation, described in 
\aalong, shows that   $X$ is 
isogenous to a product of 3 elliptic curves: 
\eqn\xisog{
X_\gamma \cong E_{\tau(\gamma) }
\times E_{\tau(\gamma)}
\times E_{\tau(\gamma) }
}
where for $\tau$ in  the upper half plane, 
\eqn\ellip{
E_\tau \equiv \IC/(\IZ + \tau \IZ)
}
and in \xisog\ $\tau(\gamma) =i\sqrt{I_4(\gamma)}$. 
The remaining vectormultiplet moduli are related to 
the values of   2-form, 4-form, 
and 6-form potentials. The remaining attractor 
equations state that these  are of type $(1,1)$ 
in the complex structure fixed by \hodgede. 
 
Let us now turn to compactifications on $X=K3 \times T^2$. 
The attractor equations for $\CN=4$ have been written in 
\adf.  Once again, the moduli must be decomposed into 
vector- and hyper- multiplets. 
Once again, we focus on the complex structure moduli.
Write $\Omega^{3,0} = \Omega^{2,0} \wedge dz$ 
where $z\  \mod (\IZ + \tau \IZ)$ is the flat coordinate 
for the $T^2$ factor of $X$. Choosing a basis for 
$H^1(T^2;\IZ)$ we can identify 
$H^3(K3 \times T^2;\IZ) \cong H^2(K3;\IZ) \oplus H^2(K3;\IZ)$
so that $\hat \gamma \cong p \oplus q$.
The attractor equations become:   
\eqn\dirsolv{
\eqalign{
2 \Im \bar C  \Omega^{2,0} = p 
\qquad & \qquad   2 \Im \bar C
\tau    \Omega^{2,0} = q  \cr}
}
for some complex constant $C$. 
This can be solved directly with the result that
 $\Omega^{2,0} =\CC( q - \bar \tau p)$
where $\CC$ is a constant, 
\eqn\solvtau{
\tau=\tau(p,q) \equiv {p\cdot q +   \sqrt{D_{p,q}} \over p^2}
}
and $D_{p,q}\equiv (p\cdot q)^2 - p^2 q^2$. 
These equations 
determine the complex structure of the  K3 surface
uniquely, by the global Torelli theorem. 
Since $\Omega^{2,0} = \CC( q - \bar \tau p)$
the Neron-Severi lattice has rank $\rho = 20$ and therefore
the transcendental lattice has rank 2. We call such 
maximally algebraic  
K3 surfaces ``attractive K3 surfaces.'' They have been completely classified
\shioda, and the classification reveals that the attractor 
variety is again related to a product of three isogenous 
elliptic curves. To explain this we need to pause and recall a few 
definitions.

An integral binary quadratic form is a matrix 
\eqn\abc{
Q = \pmatrix{ a & b/2\cr b/2 & c \cr} 
}
where $a,b,c\in \IZ$. The discriminant of the 
form is $D= -4 \det Q=b^2-4 a c$.  
The form is primitive if $g.c.d.(a,b,c)=1$, positive 
if $Q>0$ (i.e., iff $D<0, a>0$) and even if $a,c$ are even. 
To a positive quadratic form we associate an element 
$\tau$ of the 
upper halfplane via: 
\eqn\uhp{
a x^2 + b xy  + cy^2 \equiv a \vert x - \tau y \vert^2. }
Two forms $Q,Q'$ are said to be 
properly  equivalent if there is an   
$s\in SL(2,\IZ)$ such that:  
\eqn\equivfrm{
s \pmatrix{a & b/2 \cr b/2 & c \cr} s^{tr} 
= \pmatrix{a' & b'/2 \cr b'/2 & c' \cr} .
}
This induces the standard fractional linear action on $\tau$. 

Returning to K3 surfaces, if $S$ is an 
attractive  K3 surface with 
transcendental lattice $T_S$ of rank 2 then we may choose 
a $\IZ$-basis $\langle t_1, t_2 \rangle_{\IZ} = T_S$ and 
associate to it the positive even quadratic form: 
\eqn\mpsfrm{
 \pmatrix{ t_1^2& t_1 \cdot
t_2 \cr t_1 \cdot t_2 & t_2^2 \cr} 
}
Two  oriented bases for $T_S$ map to properly 
equivalent forms. By the global Torelli theorem 
the complex structure is determined by $T_S$
so there is a well-defined injective 
map from attractive 
K3 surfaces to equivalence classes 
of positive even quadratic forms. 
It turns out that the map is in fact surjective \shioda.
The proof begins with 
the torus  $A_Q = E_{\tau_1} \times E_{\tau_2}$
where 
\eqn\taus{
\tau_1 = {-b + \sqrt{D} \over  2 a} \qquad \tau_2 = {b + \sqrt{D} \over  2 } . 
} 
One first  forms the Kummer variety and then takes 
a branched double-cover. 

Applying this to our case,  to a   charge
$\hat \gamma = p\oplus q$ 
we  associate the  form of discriminant $D_{p,q}$:  
\eqn\bqform{
Q_{p,q} \equiv \half
\pmatrix{ p^2 & -p\cdot q \cr - p\cdot q & q^2 \cr}
} 
The transcendental lattice of the attractor 
K3 has a basis with Gram matrix 
$2 Q_{p,q}$. Therefore,  the attractor 
variety $X_{p,q}$ is related to a product of 
three elliptic curves: 
\eqn\threeelip{
X_{p,q} \quad {\buildrel 2:1 \over  \rightarrow}\quad   Km\Biggl( E_{\tau(p,q)} \times
E_{\tau'(p,q)}\Biggr)  \times E_{\tau(p,q)}
} 
with 
\eqn\auxtau{
\tau'(p,q)= {-p\cdot q +  \sqrt{D} \over  2 }
}

Similar considerations apply to the attractor varieties in 
the FHSV model. The double cover of the Enriques 
surface in that model is an attractive K3 surface. 
It is now determined by $p,q \in II^{10,2}(2)$. 
%
%

\newsec{Attractors are  arithmetic} 

The first connection between  arithmetic and 
supersymmetric black holes is related to  
questions about $U(\IZ)$-duality orbits of charges. 
In the $\CN=4$ compactifications the charges 
$(p,q)$ form a doublet of the $S$-duality 
group. Therefore the action of $S$-duality on 
\bqform\ is the action \equivfrm. Again we 
must pause and recall some general results. 

Quite generally, for  a fixed value of $D$ there are 
a finite number of $SL(2,\IZ)$ equivalence 
classes of forms \abc\  of discriminant $D$. 
If $Q$ is primitive, these classes are uniquely
 associated to   
ideal classes in the quadratic imaginary field 
\eqn\kaydee{
K_D \equiv \IQ[\sqrt{D}]\equiv \{ r_1 + r_2 \sqrt{D}: r_1, r_2\in \IQ \}
}
(In our case $D<0$ and $D=0,1 \mod 4$.) 
Moreover, an important  
result in number theory states
that the number of classes $h(D)$ is 
finite and grows for large $\vert D \vert$ 
like $\vert D \vert^{1/2}$. 

Returning to black holes, we apply these 
standard results to the form $Q_{p,q}$. 
Curiously, 
for large $\vert D_{p,q}\vert$, where 
the supergravity approximation is 
accurate, the near-horizon metric of 
the black hole only depends on 
$\vert D_{p,q}\vert$, and in particular the 
entropy is $S=A/4 = \pi \sqrt{-D_{p,q}}$,
where $A$ is the area of the horizon
\cvetic\fk.  A consequence is that, 
if $\langle p,q \rangle_{\IZ}\subset II^{6,22}$ is 
primitive, and if $Q_{p,q}$ is 
primitive, then the ideal classes of $K_{D_{p,q}}$ 
are in 1-1 correspondence with the $U$-duality 
classes of black holes of fixed area $A$. 
Moreover, for large $A$ the number of such 
classes grows like $A$. Pursuing this line of 
thought for other models 
leads to some interesting issues in the 
arithmetic of lattices, only briefly touched on 
in \aalong. 

The second (and more substantial) 
connection to arithmetic begins 
with the remark that if $\tau$ is quadratic 
imaginary then $E_\tau$ has the property of 
``complex multiplication.''  $E_\tau$ is a group 
so we can consider its  endomorphism algebra, 
$\End(E_\tau)$. In general this algebra is $\IZ$, 
with $n\in \IZ$ acting as 
$z \rightarrow n z $. However, 
if $\tau$ is quadratic imaginary so that 
$a \tau^2 + b \tau + c =0$, for $a,b,c\in \IZ$ then 
the  the endomorphism algebra of  
$E_\tau$ is larger than $\IZ$ because: 
\eqn\complxmult{
\omega \cdot (\IZ + \tau \IZ) \subset \IZ + \tau \IZ\qquad\quad {\rm for} \quad \qquad
\omega={D + \sqrt{D} \over  2} .
}
Under these circumstances we
say that  ``$E_\tau$ has complex multiplication by $z \rightarrow \omega z$.''

When $E_\tau$ has complex multiplication the 
 curve  has 
special arithmetic properties. Assume for simplicity that 
$j \not=0, (12)^3$. Then we can choose a Weierstrass model:  
\eqn\weiermodel{
\eqalign{
y^2 & = 4 x^3 - c(x+1) \cr
c& = {27 j \over j- (12)^3} \cr}
}
Now, one
 of the truly amazing  properties of the 
$j$ function, essentially  the first main theorem 
of complex multiplication, is that if $\tau$ is 
quadratic imaginary then $j(\tau)$ is an algebraic 
integer. Here are a few amusing examples
(Many more are available in  \cohen, and elsewhere): 
\eqn\specjay{
\eqalign{
j(i) & = (12)^3 = 1728 \cr
j(2i) & = (66)^3 = 287496\cr
j(3 i) & = 76771008 + 44330496 \sqrt{3} \cr
j(4i) & =2^3\cdot 3^3\cdot 5 \cdot 181 \cdot (210319) + 
2\cdot 3^7 \cdot 7^2 \cdot 11^2 \cdot 19 \cdot 59 \sqrt{2} \cr}
}

In fact, much more is true. The field extension
 $\widehat{K}_D\equiv K_D(j(\tau)) = 
\IQ[\tau, j(\tau)]$ is a certain Galois extension 
of the quadratic imaginary field known as a 
``class field.''  
\foot{A quite different connection between 
class field theory and string theory
and conformal 
field theory was proposed by Witten  in 
\grssmm.} 
It follows after a little work that in general
\foot{It is possible that in some examples one 
must take a further finite abelian extension of the 
class field itself.} 

{\it the attractor varieties are arithmetic varieties, 
defined over number fields associated to the field 
of definition of the periods.} 

Moreover, in the FHSV model, at an attractor 
point determined by $p_0,q_0\in II^{10,2}(2)$, 
the BPS mass-squared spectra 
are, up to an overall constant, integers reflecting 
the arithmetic of $K_D$: They are norms of 
certain ideals in the ideal class determined by 
$Q_{p_0,q_0}$. Thus, BPS mass-generating 
functions, such as occur in various quantum 
corrections to low energy effective actions, 
are generalizations of $L$-functions and $\zeta$-functions 
of $K_D$. 

\newsec{Rational theories are attractive}

The attractive K3 surfaces singled out by the 
attractor mechanism also have an interesting 
interpretation in 8-dimensional $F$-theory 
compactifications. The moduli space of the 
$F$-theory compactification is (neglecting 
the heterotic dilaton, and discrete identifications) 
a Grassmannian of spacelike 2-planes: 
\eqn\dfncrlbee{
\CB^{18,2} \equiv Gr_2^+(II^{18,2}\otimes \IR).
}
Regarding this space as a space of 
projection operators the attractor equations state that 
\eqn\leftprojs{
p_L = q_L =0 . 
}
where the subscript indicates 
the ``leftmoving projection'' onto the 18 dimensional 
negative definite  subspace of 
$\IR^{18,2}$. On the other hand, 
identifying \dfncrlbee\ as Narain 
moduli space, the equations \leftprojs\ are the 
equations defining the rational torus compactifications 
of the heterotic string, so heterotic rational 
conformal field theories (RCFT's) 
correspond to IIB compactification on 
attractor varieties. There are at least three 
interesting aspects of this connection.

First, following through $F$-theory duality one 
gets some new insight into the role of 
the Mordell-Weil group in an $F$-theory 
compactification. The torsion-free part of 
the Mordell-Weil group is an even lattice and 
therefore generates 
a chiral vertex operator algebra. This chiral 
algebra   is 
the enhanced chiral algebra of the heterotic 
compactification, and is related to the 
algebra of BPS states \hmalg. 

Second, this connection  suggests that there is a 
more general relation between heterotic RCFT 
compactifications and type II attractor 
compactifications. The realization in 8-dimensional 
compactifications is too trivial to be really 
useful. Nevertheless, an extension 
 to lower dimensional compactifications 
might be very useful. Such a connection 
realizes in a modest way part of a dream of 
Friedan and Shenker \fs:   a dense set of 
arithmetic points in  Calabi-Yau  
moduli space corresponds to rational 
conformal field theories. Indeed, the 
attractive K3 surfaces constitute an 
isolated, but dense, set of 
points in the moduli space of 
complex structures of algebraic K3 surfaces.

Third, this connection suggests an interesting 
arithmetic property of the K3 mirror map. 
The natural coordinates in heterotic compactification 
are the standard flat coordinates $y=(T,U,\vec A)$ 
parametrizing flat triples  $(G,B,\vec A)$ on $T^2$ where 
$G$ is a metric, $B$ a 2-form, and 
$\vec A$  an $E_8 \times E_8$ connection. 
The RCFT compactifications of the heterotic 
string are just those such that $y$ is quadratic 
imaginary. On the other hand, on the IIB side, 
the natural 
coordinates on \dfncrlbee\ are the $\vec \alpha, \vec \beta$ 
coefficients in the Weierstrass model of an 
elliptically fibered K3 with section \vafa\vfmr: 
\eqn\weiermodel{
\eqalign{
Z Y^2 = & 4X^3 - f_8(s,t) X Z^2 - f_{12}(s,t) Z^3 \cr
f_8(s,t) = & \alpha_{-4} s^8 + \cdots + \alpha_{+4} t^8 \cr
f_{12}(s,t) = & \beta_{-6} s^{12}+ \cdots + \beta_{+6} t^{12}\cr}
} 
The $F$-map, or $K3$ mirror map, is the map from 
an appropriate equivalence class of 
$y$ to an appropriate equivalence class  
$[(\vec \alpha, \vec \beta)]$ (the latter class 
is that following from  changes of coordinates in 
\weiermodel).  This map should have 
properties quite analogous to those of the $j$-function: 
quadratic imaginary $y$'s should map to arithmetic 
values of  $(\vec \alpha, \vec \beta)$.
In \aalong\ we verify that this is indeed the case for some 
families of K3 surfaces. 
\foot{A somewhat different arithmetic aspect of mirror 
maps was discussed by Lian and Yau in \lianyau. 
This involves the nature of the coefficients 
of mirror maps in a $q$-expansion.} 

\newsec{Attempts to understand general $\CN=2$ attractors}

It is also of great interest to extend the above 
results to more general CY compactifications. 
Naturally, 
given the examples of $T^6$ and $ K3 \times T^2$ 
one leaps to the conclusion that all attractors are 
arithmetic. This leap is formulated as three 
``arithmetic attractor conjectures'' in section 8.2 of 
\aalong. These conjectures state that at an 
attractor point the periods and the algebraic complex 
structure coordinates are both arithmetic, and the 
relation between them  generalizes the relation 
associating to quadratic imaginary 
$\tau\in K_D$ the value  $j(\tau)$ in a  class field. 
If something like the attractor conjectures 
is actually true, it would be yet another example of 
the unreasonable effectiveness of physics in 
mathematics, for it could be the beginning of a 
solution to one of the more stubborn Hilbert problems. 

Unfortunately, it is not easy to find concrete 
examples verifying the attractor conjectures.
In the context of toric constructions of 
Calabi-Yau 3-folds these conjectures are related to difficult 
questions about the arithmetic values of certain 
(GKZ) hypergeometric functions generalizing the 
 Appel and Lauricella functions.  In section 8.3 of 
\aalong\ 
we do manage to describe one slightly nontrivial 
example. It involves the two-parameter family of 
CY 3-folds of degree 8 in $\IP^{1,1,2,2,2}$ studied in 
\twop.  The attractor conjectures can 
be verified along a special divisor in the complex 
structure moduli space. Admittedly, the example does not 
really leave the world of quadratic imaginary fields
and is a bit too special to constitute major evidence 
for the attractor conjectures. Happily, one spinoff of the 
example is the demonstration that there exist 
charges with two attractor points, as mentioned above.

\newsec{Further speculations: Galois groups and height functions}

Regrettably, the connection between attractor varieties and complex multiplication  lends itself to rampant 
speculation. We have suppressed most of this but 
two further speculations do seem irresistable. 

First, in $\CN=4,8$ 
compactification since the algebraic complex structure 
moduli are valued in a class field $\widehat{K}_D$ of 
$K_D$ the Galois group ${\rm Gal}(\widehat{K}_D/K_D)$
acts on the vectormultiplet attractor moduli corresponding 
to blackholes of entropy $\pi \sqrt{-D}$ and 
permutes them. Recall that these points correspond to 
$U$-duality inequivalent backgrounds. Nevertheless, 
although $U$-duality-inequivalent, they are ``unified'' by the 
Galois action. So, does the absolute Galois group play an 
analogous unifying role for other compactifications? 

Second, the exponential growth of the dimension of 
the space of BPS states, which accounts for 
black hole entropy in the discussion of Strominger and Vafa 
\sv, is a little reminiscent of the theory of height functions. 
Since the $\CN=8$ 
attractor varieties are isogenous to products 
of elliptic curves we can explore this relation in a 
semi-quantitative way, and an attempt in this direction 
is described in \aalong. The result, which we stress 
is preliminary, is that at least for  some charges 
the black hole entropy 
$S=\pi \sqrt{I_4(\gamma)}$ compares favorably  with the 
logarithmic Faltings height $h(X_\gamma)$ of the abelian 
variety \xisog.   Using 
results of Faltings and Silverman we give a 
rough estimate in \aalong\ suggesting that  
$h(X_\gamma/\widehat{K}) \sim \kappa \log(S/\pi)$
where we expect $\kappa$ to be a simple rational 
number of order one, but could not determine it. 

\bigskip
\centerline{\bf Acknowledgements}\nobreak

I am indebted to many colleagues for important 
input. See \aalong\ for a partial list. 
I would also like to thank the Aspen Center 
for Physics and the organizers of the 
Amsterdam Summer Workshop on String Theory and 
Black Holes for hospitality during the completion 
of this paper. This work is supported by
DOE grant DE-FG02-92ER40704.

\listrefs
\bye